\documentstyle{texas}

\newcommand{\muG}{{\,\mu\rm G}}

\newcommand{\MeV}{{\,\rm MeV}}
\newcommand{\GeV}{{\,\rm GeV}}
\newcommand{\TeV}{{\,\rm TeV}}

\newcommand{\cm}{{\,\rm cm}}
\newcommand{\eV}{{\,\rm eV}}

\newcommand{\scnd}{{\,\rm s}}

\newcommand{\kpc}{{\,\rm kpc}}

\newcommand{\ster}{{\,\rm sr}}

\newcommand{\siml}
 	{\mathrel{\raise.25ex\hbox{$<$}\kern-.8em\lower.9ex\hbox{$\sim$}}}
\newcommand{\simg}
	{\mathrel{\raise.25ex\hbox{$>$}\kern-.8em\lower.9ex\hbox{$\sim$}}}
\newcommand{\mal}{{\times}}

\newcommand{\gluminosity}{[\gamma{-}{\rm luminosity}]}
\newcommand{\nuluminosity}{[\nu{-}{\rm luminosity}]}
\newcommand{\nluminosity}{[n-{\rm luminosity}]}
\newcommand{\kinematics}{\langle{\rm kinematics}\rangle}
\newcommand{\nuflux}{[\nu{-}{\rm flux}]}
\newcommand{\CRflux}{[{\rm CR-flux}]}
\newcommand{\gflux}{[\gamma{-}{\rm flux}]}
\newcommand{\opacityn}{\langle{\rm opacity}\rangle_{n}}
\newcommand{\opacityg}{\langle{\rm opacity}\rangle_{\gamma}}
\newcommand{\propagation}{\langle{\rm propagation}\rangle}

\begin{document}

\title{The relation of extragalactic cosmic ray and neutrino fluxes: the
logic of the upper bound debate}  
\author{J\"org P. Rachen$^{(1)}$, R.J. Protheroe$^{(2)}$, and
Karl Mannheim$^{(3)}$} 
\address{(1) Sterrenkundig Instituut, Universiteit Utrecht\\
Princetonplein 5, NL-3584 CC Utrecht, The Netherlands\\
}
\address{(2) Deptartment of Physics \& Mathematical Physics, The University
of Adelaide\\
Adelaide, SA 5005, Australia\\
}
\address{(3) Universit\"ats-Sternwarte G\"ottingen\\
Geismarlandstr.~11, D-37083 G\"ottingen, Germany\\
{\rm Email: J.P.Rachen@astro.uu.nl}} 

\begin{abstract}
In a recent paper, Waxman and Bahcall \cite{WB99} claimed that the present
data on ultra-high energy cosmic rays imply a model-independent upper bound
on extragalactic neutrino fluxes of $2\mal 10^{- 8} \GeV \cm^{-2} \scnd^{-1}
\ster^{-1}$, independent of neutrino energy. Mannheim, Protheroe and Rachen
\cite[v2]{MPR99} have repeated this calculation and confirmed the WB-bound,
within a factor of 2, only at a very limited energy range of $E_\nu \sim
10^{16{-}18}\eV$, while at other energies the neutrino flux is mainly limited
by the extragalactic gamma ray background to a level about two orders of
magnitude higher than the WB bound. In this paper we present a simple,
(almost) no-math approach to the problem, and discuss under which
astrophysical assumptions the WB-bound and the MPR-bound, respectively,
apply.  Then we discuss to which respect these assumptions apply to presently
discussed models of extragalactic neutrino production. We note that, averaged
over the observed luminosity function, blazars are sufficiently opaque to
ultra-high energy neutrons that there is no conflict of the predicted
neutrino fluxes with the cosmic ray data, and that these models are rather
constrained by their contributions to the extragalactic gamma ray
background. At present, no modifications are implied to the predicted
neutrino events from these models in active or planned neutrino detectors.
\end{abstract}

\section{Introduction}

The most commonly assumed mechanism for the origin of cosmic rays is the
Fermi acceleration of protons and ions at strong shock waves in magnetized
plasmas. The same mechanism also accelerates electrons, which then emit their
energy in synchrotron radiation with the magnetic field, and is therefore
probably responsible for most of the non-thermal radiation in the
universe. This has motivated theorists to search for the possible sources of
cosmic rays up to the highest energies among the known objects which emit
non-thermal radiation. A large fraction of the non-thermal power in the
universe is provided by Active Galactic Nuclei (AGN), but also the enigmatic
Gamma Ray Bursts, the most powerful explosive events in the universe, have
obtained some attention as possible sources of energetic protons, i.e.\ cosmic
rays. If cosmic rays exist in these sources, they can interact with the dense
ambient photon field, producing secondary mesons (mostly pions), which decay
and give rise to the emission of very high energy gamma-rays and
neutrinos. Moreover, such interactions can convert protons into neutrons,
which are released from the magnetic confinement and can be emitted as cosmic
rays. 

The tight connection between cosmic ray, gamma ray and neutrino production
has lead to some attempts to constrain the possible fluxes of one component
by observations of the others. This is particularly important for the
neutrino fluxes, for which large experiments are currently under
construction. Its exploration can be expected to be one of the primary goals
of the astrophysics in the next century.

\section{Extragalactic cosmic rays, gamma rays and neutrinos}

The origin of very high energy (${>}100\TeV$) neutrinos in the universe is
generally attributed to the decay of pions or other mesons (we discuss pions
only for simplicity). Pions exist in three isospin states, $\pi^+$, $\pi^-$
and $\pi^0$, which are in the gross average equally likely to be
produced. The decay of charged pions leads to the production of neutrinos and
charged leptons, while neutral pions decay electromagnetically into photons.
Under the assumption that electrons deposit their energy into photons by
radiative processes, equipartition of pion flavors leads immediately to an
approximately equal luminosity in gamma rays and neutrinos form any
pion-producing source.
\renewcommand{\theequation}{\mbox{$*$}}
\begin{equation}
\gluminosity({\rm cascade}) \approx \nuluminosity(E_\nu)
\end{equation}
One can show that this relation holds even when other mesons (e.g.\ kaons) are
involved.

The production of pions usually involves the presence or the production of
cosmic rays. In most models, pions are produced in $pp$ and $p\gamma$
interactions involving a pre-existing cosmic ray population in the source. To
specify how cosmic rays are produced, most scenarios assume Fermi
acceleration, which requires that cosmic ray protons are magnetically
confined in the source. However, confinement breaks up when a proton
interacts and changes into a neutron (isospin-flip), which can be ejected
from the source.  Charge conservation requires the production of at least one
charged meson in this interaction, and therefore inevitably leads to the
production of neutrinos. This can be expressed in the relation
\renewcommand{\theequation}{\mbox{$**$}}
\begin{equation}
\nuluminosity(E_\nu) = \kinematics \nluminosity(E_{\rm n}=f E_\nu)
\end{equation}
For both $pp$ and $p\gamma$ interactions, the factor $\kinematics$ is in the
range $0.2{-}1$, depending on the mean CMF interaction energy
\cite{sophia1,sophia2}. In some exotic models that involve the decay of
topological defects in the universe, cosmic rays and mesons are produced in
the same process and their relation can be predicted from QCD models of
hadronic fragmentation \cite{sigl}. Also this can be expressed in form of
($**$), but with $\kinematics \gg 1$.

The above relations apply to the emission process only. To establish the
corresponding relation for the {\em observable} luminosities of an
astrophysical source, we have to take into account opacity factors that may
reduce the ejected power in photons and neutrons through interactions with
ambient matter or photons. No such modification applies to high-energy
neutrinos in the cases of interest here. The interaction of neutrons
generally leads to a reduction of the bolometric cosmic ray luminosity, since
cosmic ray energy is converted into neutrinos or electromagnetic
radiation. Of particular importance is the possibility of a neutron to flip
back in such interactions into a proton, which keeps it confined over a long
time and effectively allows removing all its energy.  Photons interacting
with matter can convert their energy into heat and therefore reduce the
energy content in non-thermal radiation. The interaction of energetic non-
thermal photons with background radiation, however, does not in general lead
to a reduction of the bolometric electromagnetic energy, since the $e^\pm$
pairs emit their energy at lower frequencies in synchrotron radiation or
inverse-Compton scattering. Unless the cascade develops into saturated
comptonization, this energy can be emitted in the $\MeV{-}\GeV$ gamma ray
regime, still having non-thermal characteristics.

To infer relations between observable fluxes, a further factor arising from
the different propagation properties of gamma rays, neutrinos and cosmic rays
has to be considered. Neutrinos, and gamma rays ${\ll}100\GeV$, are not
strongly affected by the presence of cosmic photon or matter backgrounds.
The luminosity flux relation is therefore given by standard cosmological
expressions in both cases. Cosmic rays ${\simg}10^{18}\eV$ are affected by
photohadronic pair and pion production with the local cosmic microwave
background. Also at lower energies these interactions can modify the relation
between cosmic ray and neutrino fluxes arriving from large redshifts. In
addition, cosmic ray propagation can be influenced by the presence of
magnetic fields in the source, our galaxy and on large scales.
\setcounter{equation}{0}
\renewcommand{\theequation}{\mbox{\bf relation \arabic{equation}}} 

Following this, we may write the relation of observable neutrino fluxes
to observable and gamma ray and cosmic ray fluxes as
\begin{equation}
\nuflux({\rm bolometric}) < \gflux(\MeV{-}\GeV) \opacityg
\end{equation}
and
\begin{eqnarray}
\nuflux(E_\nu) &\le& \CRflux(E_{\rm CR} = f E_\nu) \\
	& & \times \opacityn \kinematics \propagation \nonumber 
\end{eqnarray}
where the value of $\propagation$ is determined by the inverse ratio of the
proton attenuation length to the Hubble radius, and by the dependence of the
comoving source luminosity density on redshift. For non-evolving sources, it
ranges from ${\approx} 1$ for $E_{\rm CR}<10^{17}\eV$, over ${\sim} 3$ at
$E_{\rm CR} = 10^{19}\eV$ to ${\simg} 100$ for $E_{\rm CR} >
10^{20}\eV$. With source evolution on the level suggested by observations for
AGN and starburst galaxies, the value at $E_{\rm CR} \simg 10^{19}\eV$ is
increased by about a factor of $5$. The parameter $f \approx 0.01{-}0.05$ is
determined from the interaction kinematics and depends on the CMF energy
\cite{sophia1,sophia2}.

Relation 1 holds for the comparison of bolometric luminosities, and considers
that a (maybe dominant) fraction of the non-thermal radiation of the source
is not produced in photohadronic interactions, but by synchrotron-self
Compton emission of co-accelerated primary electrons. The factor $\opacityg$
considers the fraction of energy reprocessed to energies below $\sim\MeV$ by
comptonization. In relation 2, which holds also for specific energies, we
have considered the possibility that part of the cosmic ray ejection is due
to the direct, non-adiabatic ejection of protons.

\section{Can we impose ``robust'' or ``model-independent'' upper bounds?}

Obviously, both relations 1 and 2 pose upper limits on the observable
neutrino flux, if we chose proper upper limits for the parameters averaged
over all contributing sources.  $\gflux$ and $\CRflux$ on the right hand
sides are observables, so it is in principle possible to fix their values by
observations. All other parameters have to be determined by theory. We may
distinguish, however, whether the theory used here is well established and
supported sufficiently by observations, or whether we have to rely on weakly
supported hypotheses. For example, the parameter $\kinematics$ is, within
narrow bounds, well known and little dependent on astrophysical model
assumptions, as long we confine ourselves to neutrino production in $p\gamma$
or $pp$ interactions. More difficult is the situation for $\propagation$ ---
while its value is well known for a given cosmic ray energy assuming
straight-line propagation, since it only depends on well measured cross
sections and the temperature of the microwave background, the influence of
poorly known magnetic fields in the universe is difficult to determine.
Obviously, the $\opacityg$ and $\opacityn$ can hardly been constrained a
priory without specifying a particular choice of sources.

As an additional complication, also the determination of $\gflux$ and
$\CRflux$ is not straightforward, since we have to distinguish the {\em
extragalactic} contribution to these fluxes from the generally dominant
galactic contribution. However, here we may remember the logic inherent to an
``upper bound'': we do not need precise measurements for these quantities, it
is sufficient to have upper limits for them. Obviously, a safe upper limit
for any extragalactic flux contribution is, unless determined more precisely,
given by the total observed flux.  The limits determined by this minimal
condition, however, may be very weak and of little practical relevance.

The extragalactic $\MeV{-}\GeV$ gamma-ray background (hereafter EGRB%
\footnote{For the convenience of a simple argument, we understand in this
paper under EGRB the diffuse {\em plus} identified point source contribution
to the extragalactic gamma ray background, thus the total density of ambient
extragalactic gamma rays. We note that usually given experimental values for
this quantities subtract extragalactic point sources.}
) is fairly well determined \cite{EGRB}.  Relation 1 has therefore readily
been used to normalize flux estimates for extragalactic neutrinos. A
complication is only that there is little theoretical agreement in which
energy range exactly the reprocessed electromagnetic cascade radiation
emerges from the source. For example, normalizing to the EGRB above $1\MeV$
yields about one order of magnitude higher fluxes than normalizing to the
EGRB above $100\MeV$.

The extragalactic contribution to the cosmic ray spectrum is generally
believed to dominate the total observed flux above $3\mal 10^{18}\eV$, where
the cosmic ray spectrum shows a distinct feature, called ``the ankle''. This
belief is also somewhat supported by the absence of a signature of the
galactic plane in the arrival direction distribution of the cosmic ray
events, but it may be noted that models for a galactic halo origin of cosmic
rays, which would not show such a signature, have been suggested.  Another
clear signature of an extragalactic origin would be the
Greisen-Zatsepin-Kuzmin (GZK) cutoff expected above $\sim 5\mal 10^{19}\eV$
due to photoproduction losses of the cosmic rays at the microwave
background. Unfortunately, the current experiments disagree about the
presence or absence of this cutoff in the data, so that we have to wait for a
better data statistics to achieve clarification.

Below the ankle, very little is known about the origin of the dominant cosmic
ray component. Some experiments suggest a composition change from a
dominantly heavy (iron) component to a light (proton) component at the ankle,
i.e., at the same position where also the spectrum flattens.  The result was
recently corroborated by measurements at lower energies, which indicate an
increasingly heavy composition of cosmic rays around ``the knee'' of the
CR-spectrum at $10^{14{-}16}\eV$. This, and also some tentative results on a
possible signature of the galactic plane in the arrival directions at
$10^{17}\eV$, has lead to a common sense that cosmic rays below the ankle are
dominantly of galactic origin.  Although this might be the case, and it is
indeed expected from theoretical arguments, too, we have to be careful not to
over-interpret these data. In particular the composition measurements at
$10^{17}\eV$ rely on Monte Carlo simulations using particle interaction cross
section extrapolations into energy regions which are not explored by
accelerators yet. It has been shown that the presence or absence of the
composition change signature in different experiments is dependent on the
Monte-Carlo code used for the data interpretation \cite{dawson}.  In summary,
we may have good observational evidence that somewhere between ``the knee''
and ``the ankle'' of the cosmic ray spectrum the dominant origin changes from
galactic to extragalactic. Any stronger statement on the exact shape of the
extragalactic spectrum below the ankle, however, rather falls into the
category of personal or public opinion. The only conservative upper limit on
the extragalactic cosmic ray flux below the ankle and above the knee is
therefore the measured total flux.

\section{The Waxman-Bahcall upper bound}

In a recent paper, Waxman and Bahcall \cite{WB99} claimed that the relation
between cosmic ray and neutrino fluxes (relation 2) sets a model-independent
upper bound on extragalactic neutrino fluxes at all energies. This bound is
about $1{-}2$ orders of magnitude stricter than previously assumed bounds
from comparison with the EGRB (relation 1). Consequently, the authors claim
to rule out most present models of neutrino production, in particular those
connected to hadronic AGN models that have been generally normalized to the
EGRB. As a corollary, they claim that this provides a model-independent proof
that the EGRB is not completely produced by hadronic processes in AGN.

From the discussion above, this conclusion seems rather surprising,
particularly regarding to the fact that their bound is energy independent in
power per decade---a behavior that is not seen in the cosmic ray data, which
they claim are the only pinpoint for their conclusion. It is therefore worth
asking: (a) how exactly this bound was derived; (b) in which respect it is
really model independent, which means that it affects any model suggested for
extragalactic neutrino production in the past, present and future; (c) to
which respect it really affects present models for neutrino production in AGN
jets; and (d) whether it really rules out a hadronic production of the EGRB.

Indeed, Waxman and Bahcall (WB) use relation 2 to derive their bound,
although this is somewhat hidden.  Instead of writing down the relation of
neutrino and cosmic ray mean free paths, they use a result obtained in an
earlier paper for the local cosmic ray injection density at $10^{19}\eV$,
where the propagation of ultra-high energy (UHE) cosmic rays was properly
treated using a common transport approximation \cite{Wax95b}. Applying then
the trivial equations for cosmological neutrino transport, they derive the
correct (straight-line) $\propagation$ factor used in relation 2. They also
discuss properly the dependence of this factor on source evolution. Also
their factor $\kinematics = 0.25$ falls into the right range for
photohadronic (or $pp$) neutrino production. Also, WB point out that no
statement can be made on sources which are opaque to cosmic ray neutrons,
which they exclude a priori from their treatment.

We may at this point summarize the assumptions that so far entered in the
derivation of the WB-bound:
\begin{description}
\item[Assumption 1:] Neutrinos are produced in interactions of cosmic rays
with background photons or matter.
\item[Assumption 2:] The sources are transparent for neutrons of an energy
${\sim} 10^{19}\eV$.
\item[Assumption 3:] Cosmic rays of energy $10^{19}\eV$ ejected by these
sources are not affected by magnetic fields in or at the vicinity of the
source, or on large scales.
\end{description}
While Assumption 2 is clearly stated in the paper, assumption 1 is rather
implicitly understood. It is certainly justified since in fact most models of
extragalactic neutrino production invoke this process.  In contrast, WB
devote an extensive discussion to the justification of assumption 3, where
they show that (a) neutrons of this energy ($10^{19}\eV$) escape from most
known strong field regions around putative neutrino sources before undergoing
$\beta$-decay, and (b) protons of this energy cannot be confined in large
scale fields (i.e.\ clusters of galaxies or superclusters) for time scales
comparable to the Hubble time. From this, they conclude that magnetic fields
cannot lead to inhomogeneities in the universal distribution of cosmic rays,
which can be easily seen to be equivalent to the straight-line propagation
assumption. Thus, according to WB, assumption 3 can be derived from our
present observational upper limits on extragalactic magnetic fields.  The
authors neglect, however, that particles moving diffusively through an
adiabatically decreasing magnetic field suffer energy losses due to expansion
work towards the outer medium. We return to this issue below.

Under assumptions 1-3, the bound is only valid at one energy of the spectrum:
it corresponds to a cosmic ray energy of $10^{19}\eV$, or a neutrino energy
of $\sim 3\mal10^{17}\eV$ assuming standard kinematical relations.  To extend
it to other energies, WB introduce
\begin{description}
\item[Assumption 4:] The overall cosmic ray injection spectrum in the
universe has the spectral shape $dN/dE \propto E^{-2}$, and extends without break
up to $10^{19}\eV$ and beyond.
\end{description}      
To support this assumption, WB refer to the theory of diffusive shock
acceleration, which canonically predicts an $E^{-2}$ power law spectrum for
particle acceleration at strong, non-relativistic shocks. They give no reason
why the spectrum should extend to $10^{19}\eV$ for all sources in the
universe.

With assumption 4, the spectral shape of the WB bound becomes obvious: since
the factor $\kinematics$ is only weakly dependent on energy (see M\"ucke et
al. \cite{sophia2}, these proceedings, for a more detailed discussion), the
assumed flat (= constant power per decade) cosmic ray spectrum produces a
flat neutrino spectrum. {\em The WB upper bound is therefore the result of
fitting a model spectrum to the observed cosmic ray flux at $10^{19}\eV$.}

\section{Critique of the Waxman-Bahcall bound as a general upper limit}

Obviously, Waxman and Bahcall made no ``mistake'' in deriving their bound ---
for extragalactic neutrino sources which comply with assumptions 1-4, it is
indeed a valid upper limit for the observable flux, based on
observations. The question we have to ask is whether this justifies the
claim of ``model-independence'': is it in fact reasonable to believe that
assumptions 1-4 are all of general validity for any possible neutrino source?

Assumption 1 is certainly the one that can most easily be accepted; it is the
only mechanism predicting high-energy neutrinos so far which is not entirely
speculative.  Nevertheless, we may remind the reader that other models have
been suggested. Such models, which are based on string hadronization (the
so-called ``topological defect'' models for cosmic ray origin \cite{sigl}),
predict for the given cosmic ray flux a neutrino flux about two orders of
magnitude larger than the WB bound. This is based a relation similar to our
relation 2, but with $\kinematics \sim 100$, rather than $0.25$ as used by
WB. It has been shown that these models are strongly constrained by the more
general relation 1, i.e.\ the requirement not to over-produce the EGRB
\cite{PJ96}.

Regarding assumption 2, we may just follow Waxman and Bahcall and restrict
our consideration a priori to sources fulfilling it. However, we disagree
with WB in stating that such sources cannot be identified or constrained by
any other emission than neutrinos. For a large subclass of them, i.e., those
who emit gamma rays in the $\MeV{-}\GeV$ (but not up to the $\TeV$) regime,
non-thermal gamma emission produced from $\pi^0$-induced unsaturated
synchrotron-pair cascades can emerge from the source. As correctly noted by
Waxman and Bahcall, there is a strict connection between $\gamma\gamma$ and
$n\gamma$ opacity \cite{WB99,MPR99}. While sources transparent to $\TeV$
gamma rays can be shown to be transparent to UHE neutrons, too, one can use
the same relation to show that sources which show an opacity break in the
gamma ray spectrum at a few $\GeV$ must be opaque to neutrons at
$10^{19}\eV$. For example, this is the case for most high luminosity gamma
ray blazars. In fact, the observed non-thermal gamma-ray emission from such
blazars did motivate the assumption that they are strong neutrino sources,
and at the same time it restricts the maximum neutrino flux from such sources
by relation 1 to a level about two orders of magnitude above the WB bound. We
discuss below in which respect the WB bound still affects the expected
neutrino fluxes from blazars.

Before discussing assumption 3, we turn to assumption 4, which is certainly
the one with the weakest observational support. In fact, even the theoretical
support WB give has to be questioned. For the large variety of shocks of
different speeds and compression ratios in astrophysical sources, shock
acceleration theory predicts a large range of spectral indices; the common
value of $2$ is just as a canonical assumption, applying to non-relativistic,
strong shocks. Even more questionable is the assumption that in all sources
the spectrum extends to $10^{19}\eV$. Obviously, even if all accelerators
would have a spectral index of $2$, the contribution of sources with cutoffs
below $10^{19}\eV$ can locally produce an overall spectrum much steeper.
This would allow higher associated neutrino fluxes at energies below
$10^{17}\eV$, without being in conflict with the cosmic ray data at
$10^{19}\eV$ where only a few sources contribute. Obviously, the existence of
Fermi accelerating sources with a proton maximum energy below $10^{19}\eV$
cannot be ruled out from first principles. Rather, it is suggested by a
consistent interpretation of gamma ray observations of blazars within the
hadronic model.

Without assumption 4, which allows to restrict the comparison of neutrino and
cosmic ray transport properties to the energy $10^{19}\eV$, the validity of
assumption 3 has to be revised. Taking the results of WB for confinement in
cores of rich galaxy clusters and large scale supergalactic filaments, we
obtain confinement over a Hubble time for protons below $10^{18}\eV$ and
$10^{16}\eV$, respectively. We could apply the same calculation to galactic
halos of magnetic field strength $1\muG$ on a variation scale of $10\kpc$,
extending to ${\sim} 300\kpc$, and again obtain confinement for cosmic rays
below $10^{16}\eV$. At the same energy, we can also expect the halo of our
own galaxy to modify the incoming extragalactic proton flux, similar to the
modifications of the solar wind observed in cosmic rays around $1\GeV$. Since
${\sim} 10^{16}\eV$ protons are connected to the production of ${\sim}
300\TeV$ neutrinos, it would be unreasonable to propose an upper bound on
their extragalactic flux based on cosmic ray observations {\em directly
related to their energy}, regardless what the limits on the extragalactic
contribution to the observed cosmic ray flux at ${\sim} 10^{16}\eV$ are.

Moreover, protons which migrate through a gradually decreasing magnetic
field, with a gyro-radius much smaller that the scale over which the field
changes, lose energy towards adiabatic expansion if there is some interaction
with the cold plasma which allows them to keep an isotropic distribution in
the rest frame of the bulk flow. The detailed energy loss depends on the
exact field configurations, but for the simplest case of a largely chaotic
field, the energy loss follows the same rule as the adiabatic expansion of a
relativistic gas, i.e.\ $E_{\rm CR} \propto R^{-1}$. If this applies to the
putative outflows in galactic halos (galactic winds), and if we assume that
most neutrino sources reside in galaxies having winds, then the cosmic ray
bound can be relaxed by one order of magnitude or more for neutrino energies
below $10^{17}\eV$. A similar effect can be obtained by the likely assumption
that a considerable fraction of neutrino sources reside in galaxies belonging
to more or less dense clusters or groups with stronger-than-average magnetic
fields between galaxies, leading to a (partial) large scale confinement of
cosmic rays below $10^{18}\eV$.

We note, in agreement with Waxman and Bahcall, that none of the above effects
can strongly influence the propagation of cosmic rays at ${\simg}
10^{19}\eV$. One reason for this is, that neutrons of this energy jump out of
the confinement of most of the structures we discuss before undergoing
$\beta$-decay. Our critique is rather directed against the connection of the
justification of assumption 3 on the validity of assumption 4. Dropping
assumption 4, i.e.\ the application of a model spectrum, the influence of
magnetic fields can no longer be neglected. On the basis of relation 2,
together with assumptions 1 and 2, we have therefore derived a neutrino upper
bound which is truly based on the observed cosmic ray flux \cite{MPR99}. We
have also discussed the possible influence of magnetic fields on this
bound. The result is that, as expected, we confirm the WB bound for a
neutrino energy of $\sim 3\mal 10^{17}\eV$, but find much less restrictive
limits at lower energies. At neutrino energies below about $10^{15}\eV$, the
flux is only limited by the EGRB, regardless of the choice of parameters.

Both, the cosmic ray data we use and the magnetic fields we assume, suffer
from difficulties in the interpretation of the data, and can therefore be
disputed. However, at this point we may remind again in the logical meaning
of the term ``upper bound'': to derive a true upper bound, we can only use
the observational {\em upper limits} on both the extragalactic cosmic ray
flux and the magnetic fields connected to extragalactic sources and large
scales. Everything else would not comply with the standards of a reliable
scientific result. It is needless to say that we do not want to {\em propose}
neutrino fluxes of this strength --- we only state that they cannot be ruled
out by general theoretical arguments and current observations. It is also
worth to note that the qualitative feature of our result, i.e., that our
bound is nowhere as strict as at $3\mal 10^{19}\eV$, is independent whether
we use or don't use the cosmic ray composition data, or whether we assume or
don't assume an effect of magnetic fields. 
 
At last, we may also have a look on neutrino energies higher than $3\mal
10^{17}\eV$, which are produced by cosmic rays above $10^{19}\eV$. Here, as
we see immediately from relation 2, the factor $\propagation$ rises from $3$
to about $100$ (for the no-evolution case). Assuming that the observed
quantity $\CRflux$ does not drastically change, this would imply a strong
increase of the upper bound. In fact, when we look at the data, a
continuation of the cosmic ray spectrum as a power law $dN/dE\propto
E^{-2.7}$ beyond $10^{20}\eV$ is suggested by one of three large exposure
experiments \cite{AGASA}, and consistent with the combined result of all
experiments (including the ones with lower exposure), and can therefore not
be ruled out. The fact that the WB bound does not show this increase again
goes back on assumption 4: The assumption of a flat injection spectrum {\em
implies} a drastical change in the slope of the observed CR spectrum,
i.e.\ the existence of the GZK-cutoff. This is also consistent with present
data, but we note that it is the current {\em lower} limit on the observed CR
flux at this energy, and can therefore not be used for upper limit
estimates. We also point out that the common assumption that the post-GZK
cosmic rays origin from a strong, local source, would {\em not} imply a
increase of the neutrino flux: For a local source, the factor $\propagation$
in relation 2 obviously approaches unity, since both cosmic rays and
neutrinos propagate (approximately) loss-free and in straight lines.  An
increase of the neutrino flux correlated to our bound would imply that the
non-observation of the GZK cutoff is due to the increased activity of all CR
sources in the universe, rather than to one local source. Although this
scenario is currently not favored by theoretical arguments, it cannot be
ruled out a priory. Only an {\em observational} upper limit excluding the
associated neutrino flux would so this. We note that the present theoretical
upper limit on the UHE neutrino flux is set by the observed EGRB through
relation 1.

\section{The impact of the Waxman-Bahcall upper bound on present models}

In the last section we have shown how the special selection of parameters and
assumptions allowed Waxman and Bahcall to set their cosmic ray upper bound to
the lowest value possible for any model assumption. We have also shown that
there is a large freedom to invent models which evade this bound at energies
different from those chosen by WB in their derivation. Here we want to
discuss in which respect their bound affects present models which are already
discussed in the literature. Clearly, since such models make a clear
prediction about the global source spectrum, the bound may be of more
relevance here since we can compare in the most restrictive energy regime,
$E_{\rm CR} \approx 10^{19}\eV$ or $E_\nu\sim 3\mal 10^{17}\eV$.

We start with AGN models. Waxman and Bahcall have already noted that there is
one class of AGN related neutrino models for which no bound whatsoever can be
stated, except by direct neutrino observations: the so called AGN core model
\cite{SDSS92}, which is opaque to both neutrons and gamma rays. (Actually,
there is a bound also on this models, since the energy in gamma rays is
converted to X-rays by saturated comptonization, and can be compared to the
total flux of observed extragalactic X-ray point sources and the diffuse
background. This is the way how this model indeed has been normalized.) Here
we concentrate, like Waxman and Bahcall, on the discussion of AGN jet models.

First of all, it should be noted that Waxman and Bahcall did not discover
that such models may be constrained by cosmic ray data. Mannheim (1995,
\cite{Man95}) already pointed out this problem, and suggested two models:
Model A, which was constructed to explain both the cosmic ray data (assuming
neutron transparence and straight-line propagation) and the EGRB above
$100\MeV$ (however, using an incorrect relation of gamma ray and neutrino
fluxes, see M\"ucke et al. \cite{sophia2}, these proceedings), and Model B,
normalized to the gamma ray background above $1\MeV$, which was at that time
overestimated by the incorrect Apollo measurements by one order of magnitude.
It was noted in ref.~\cite{Man95} that for Model B, in order to evade
overproduction of cosmic rays above the ankle, one has to assume some
mechanism preventing their cosmic rays from reaching us at these energies.

Citing Model B only, and two similar models by Protheroe \cite{Pro97}, and
Halzen and Zas \cite{HZ97}, Waxman and Bahcall claimed that all these models
violate the cosmic ray bound by two orders of magnitude and can therefore be
ruled out. In fact, using the correct bound derived for the cosmological
evolution observed in AGN, the discrepancy is reduced to a factor ${\sim} 30$,
and by another factor of $2$ when we recalculate the WB bound at $E_{\rm CR}
= 10^{19}\eV$ using a precise Monte-Carlo simulation of extragalactic
transport \cite{PJ96}, rather than the approximate treatment performed by
Waxman (1995, \cite{Wax95b}).

Obviously, hadronic blazar models follow assumption 1, and since we have a
model spectrum given we can chose an energy in the spectrum where assumption
3 applies as well. To support the validity of assumption 2 (transparence) for
AGN jets, WB refer to the observed TeV emission of Mrk 421 and Mrk
501. Unfortunately, they misinterpret the TeV data in stating that the
observed emission at $10\TeV$ proves that blazar jets are optically thin at
this energy. In fact, the observed break of the gamma ray spectral index
between the EGRET regime (${<}30\GeV$) and the Whipple/HEGRA data
(${>}300\GeV$) \cite{SBB98} implies that these sources become optically thick
at ${\siml}300\GeV$. It should be pointed out that in an homogeneous emitter,
a $\gamma\gamma$-opacity larger than one does not lead to an exponential
cutoff, as sometimes erroneously assumed, but to a spectral break by the
amount of the low energy flux spectral index. For Mrk\,421 and Mrk\,501, the
observed $\GeV{-}\TeV$ break matches this prediction very well. Moreover, a
spectral break of this kind is not expected in the emission spectrum of the
hadronic scenario, so a consistent interpretation of the data within this
model requires the assumption that the observed break is due to
opacity. Therefore, the $\gamma\gamma$-opacity of Mrk 501 at 10 TeV is
${\sim} 30$, rather than ${\ll} 1$ as assumed by WB. Correcting this in the
estimate, we obtain an neutron-opacity of Mrk\,501 at $10^{19}\eV$ of ${\sim}
0.1$. Mrk\,501 is therefore indeed optically thin for neutrons, but we have
to note that it is a low-luminosity blazar, and that the opacity is directly
proportional to the blazar luminosity. Averaging the neutron opacity of
blazars over the flat blazar luminosity function \cite{wolter}, we obtain an
average value $\opacityn \sim 10$, which reduces the cosmic ray ejection per
given neutrino flux (or, in other words, increases the bound) by the same
factor (see \cite[v2]{MPR99} for details). This removes the discrepancy with
the WB bound --- existing AGN jet models are {\em not at any energy} in
conflict with the cosmic ray data, because they do not fulfill assumption
2. Obviously, this also shows that a hadronic production of the EGRB is not
ruled out by the cosmic ray constraint, as Waxman \& Bahcall claim. Rather,
improved determinations of the EGRB and its origin may set the strongest
constraints on the possible neutrino fluxes. We note that this result was
obtained without any modifications to the models, and without invoking other
energy loss processes for cosmic rays expected in the extended halos of
radio-loud AGN.

As a side remark, we note that due to the very flat neutrino spectrum AGN
models expect (and always expected) neutrino fluxes in the interesting PeV
regime which are much below the WB bound. Even if the bound would be
perfectly valid, a direct discrepancy at these energies never existed.

We now may add a few remarks on neutrinos from Gamma Ray Bursts. Here, no
discrepancy with the cosmic ray bound has been found by Waxman and
Bahcall. We may remark, that due to the property of this model to expect an
optically thin, $E^{-2}$ cosmic ray emission spectrum, and cutoffs generally
above $10^{19}\eV$, it is the only model to which all assumptions of Waxman
and Bahcall, thus also their bound, fully apply. However, we may point out an
interesting turn of the argument: In highly relativistic flows like GRB (and
AGN also), we have good reason to assume that the direct ejection of protons
is strongly suppressed, because the adiabatic loss time is of the order of
the crossing time of the shell. Thus, it is likely that only neutrons can be
ejected from a GRB shell \cite{RM97}. If this is the case, then we can use
the WB bound for an $E^{-2}$ neutron spectrum, as expected to be produced by
GRB, as a test flux for the hypothesis that GRB are the dominant sources of
ultra-high energy cosmic rays \cite{Wax95}. If it could be independently
confirmed that GRBs follow the evolution pattern observed in star-formation,
an observed neutrino flux correlated with GRB events ``only'' on the level
predicted by Waxman and Bahcall (1997 \cite{WB97}), which is then about one
order of magnitude below the appropriate bound, would provide evidence {\em
against} rather than in favor of this scenario.

\section{Conclusions}

The relation of cosmic ray and neutrino fluxes has been shown to be an
important, so far not sufficiently noticed measure for the viability of
models of extragalactic neutrino production. It is thanks to Waxman and
Bahcall that this point has now found attention by the
community. Unfortunately, the way their result was presented, namely as a
model-independent upper bound on any kind of extragalactic neutrino
production, could impose to some severe misunderstandings. The most serious
could be that this result might shatter our confidence in the object of
very-high and ultra-high energy neutrino observatories.

In this paper we have presented the case that the Waxman-Bahcall upper bound
is not model-independent, but rather relies on very special model
assumptions. We have also shown that present models for extragalactic
neutrino fluxes, which have provided one motivation for the construction of
the experiments mentioned above, are not seriously affected by their result
and need no modifications. However, it is also clear that the consistency of
these models with cosmic ray observations is marginal, so that cosmic ray data
can be regarded an important constraint for their parameter space.

With respect to the motivation of experiments, we may make one point very
clear: The debate whether the Waxman-Bahcall bound is valid or not is a
purely theoretical dispute. It is the dispute whether assumptions 1-4 stated
in Section 3 generally apply to nature, or whether they do not. The decision
can only be made by experiment. Although theories are necessary to understand
our data, they can never replace them. Truly model-independent bounds are
only {\em observational} upper limits.

The discussion in this paper also made clear how many important questions
regarding the origin of cosmic rays can be decided by neutrino
observations. The prediction of neutrinos above the WB-bound at energies
$10^{16}-10^{18}\eV$ is an important test of the viability of hadronic blazar
models, and of the total contribution of hadronic blazar emission to the
extragalactic gamma ray background. Setting upper limits to neutrinos from
Gamma-Ray Burst below the WB-bound may enable us to limit their contribution
to the ultra-high energy cosmic ray spectrum --- or, finding them on the
level of the bound, would provide strong evidence that they are indeed the
dominant sources of these cosmic rays. Finally, in case that the
non-existence of the GZK cutoff in the cosmic ray spectrum is further
supported by observations, searching for neutrinos in excess of the WB-bound
at ultra-high energies (${>}10^{19}\eV$) can test whether this is due to an
increased overall activity of the cosmic ray/neutrino sources in the
universe, or rather due to the contribution from one, local source (or even
our own galactic halo). All in all, these are only a few reasons why the
tight connection between extragalactic cosmic ray and neutrino fluxes
provides a strong additional motivation for VHE and UHE neutrino
observatories.

\section*{Acknowledgements}

JPR acknowledges support by the EU-TMR network ``Astro-Plasma Physics'' under
contract number ERBFMRX-CT98-0168. RJP is supported by the Australian
Research Council.

\end{document}